\documentclass[aps,prl,twocolumn,groupedaddress,showpacs]{revtex4}

\usepackage{bm}
\usepackage{graphicx}

\begin{document}

\title{Evolution from incoherent to coherent electronic states and its implications to superconductivity in FeTe$_{1-{\it x}}$Se$_{\it x}$}

\author{E. Ieki,$^1$ K. Nakayama,$^1$ Y. Miyata,$^1$ T. Sato,$^1$ H. Miao,$^2$ N. Xu,$^{2}$ X.-P. Wang,$^2$ \\ P. Zhang,$^2$ T. Qian,$^2$ P. Richard,$^{2,3}$ Z.-J. Xu,$^4$ J. S. Wen,$^4$ G. D. Gu,$^4$ H. Q. Luo,$^2$ H.-H. Wen,$^5$ H. Ding,$^{2,3}$ and T. Takahashi$^{1,6}$}

\affiliation{$^1$Department of Physics, Tohoku University, Sendai 980-8578, Japan\\
$^2$Beijing National Laboratory for Condensed Matter Physics, and Institute of Physics, Chinese Academy of Sciences, Beijing 100190, China \\
$^3$Collaborative Innovation Center of Quantum Matter, Beijing, China\\
$^4$Condensed Matter Physics and Materials Science Department, Brookhaven National Laboratory, Upton, New York 11973, USA \\
$^5$Center for Superconducting Physics and Materials, National Laboratory for Solid State Microstructures, Department of Physics, Nanjing University, Nanjing 210093, China \\
$^6$WPI Research Center, Advanced Institute for Materials Research, Tohoku University, Sendai 980-8577, Japan
}

\date{\today}

\begin{abstract}
We have performed systematic angle-resolved photoemission spectroscopy (ARPES) of iron-chalcogenide superconductor FeTe$_{1-{\it x}}$Se$_{{\it x}}$ ($0 \leq {\it x} \leq 0.45$) to elucidate the electronic states relevant to the superconductivity. While the Fermi-surface shape is nearly independent of {\it x}, we found that the ARPES spectral line shape shows prominent {\it x} dependence. A broad ARPES spectrum characterized by a small quasiparticle weight at ${\it x} = 0$, indicative of incoherent electronic states, becomes progressively sharper with increasing {\it x}, and a well-defined quasiparticle peak appears around ${\it x} = 0.45$ where bulk superconductivity is realized. The present result suggests the evolution from incoherent to coherent electronic states and its close relationship to the emergence of superconductivity.
\end{abstract}


\maketitle
The discovery of superconductivity at 26 K in LaFeAsO$_{1-{\it x}}$F$_{\it x}$ \cite{Kamihara} and the subsequent increase of superconducting transition temperature ($\it{T}_{c}$) up to $\sim$55 K \cite{Ren, Kito, Wang} have evoked much attention to this new family of high-$\it{T}_{c}$ materials called Fe-superconductors. The recent report on a possible higher-$\it{T}_{c}$ superconductivity at $\sim$65 K in a FeSe thin film \cite{Xue, Zhou} has brought a further excitement and generated tremendous attention to this new class of Fe-superconductors, iron chalcogenide FeTe$_{1-{\it x}}$Se$_{\it x}$ \cite{Hsu, Yeh, Fang, Mizuguchi, LiuNatureMat}. From the crystallographic point of view, FeTe$_{1-{\it x}}$Se$_{\it x}$ is regarded as the simplest Fe-superconductor, since it consists of only Fe(Te,Se)-tetrahedra layers. It is also known that the end material, FeTe (${\it x} = 0$), exhibits a ``bicollinear" antiferromagnetic order with the propagation vector of $Q =(\pi/2,\pi/2)$, which differs from that of the pnictide family exhibiting the ``collinear" antiferromagnetic order with $Q = (\pi, 0)$ \cite{LiuNatureMat, Li}. Another interesting aspect of FeTe$_{1-{\it x}}$Se$_{\it x}$ is that the doping of extra carriers is not necessary to induce superconductivity. The isovalent substitution of Se for Te, instead of carrier doping, suppresses the bicollinear antiferromagnetic order for ${\it x}  > 0.1$, and bulk superconductivity takes place around ${\it x} = 0.3$ [see Fig. 1(a)] \cite{LiuNatureMat}. Upon further substitution, ${\it T}_{\it c}$ reaches the maximum of $\sim$14 K around ${\it x}  = 0.5$ and then decreases to 8 K at the end composition of ${\it x}  = 1$ (FeSe).

\begin{figure*}
\includegraphics[width=6.4in]{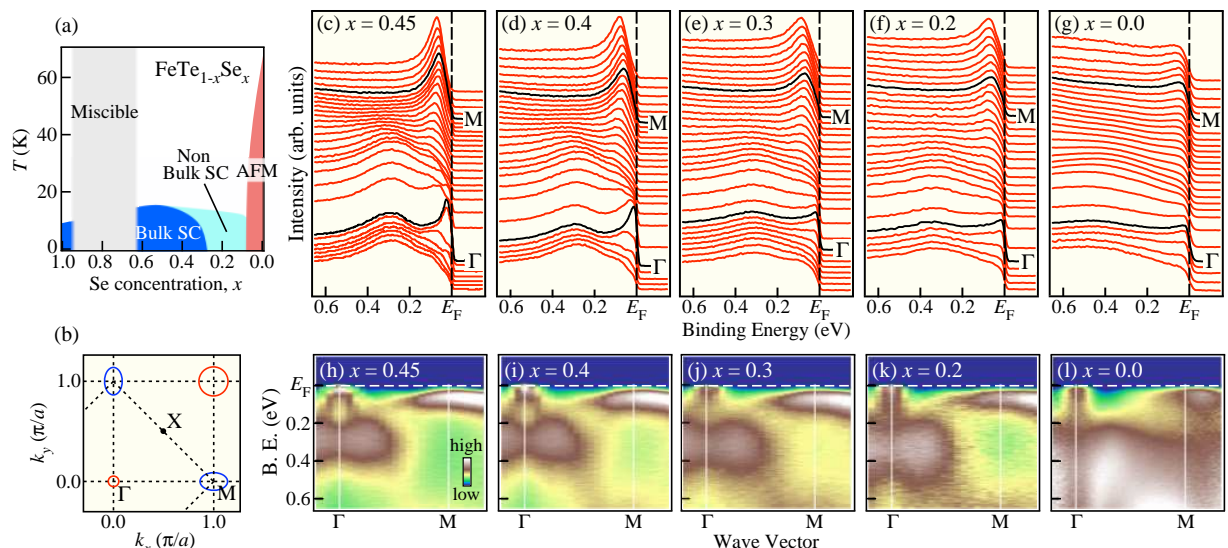}
\vspace{-0.5cm}
\caption{(Color online) (a) Schematic phase diagram of FeTe$_{1-{\it x}}$Se$_{\it x}$ derived from refs. 9, 11, and 13. SC and AFM denote superconductivity and antiferromagnetism, respectively. The miscible region exists at high Se concentrations due to the difficulty in growing single-phase samples. (b) One-Fe/unit-cell Brillouin zone of FeTe$_{1-{\it x}}$Se$_{\it x}$ used in this study together with the schematic hole and electron Fermi surfaces at the $\Gamma$ and M points, respectively. (c)-(g) Se-concentration dependence of normal-state ARPES spectra along the $\Gamma$-M line in a wide energy region for FeTe$_{1-{\it x}}$Se$_{\it x}$ (${\it T} = 25$ K for ${\it x}  = 0.45$-$0.2$ and 80 K for ${\it x} = 0.0$) measured with the He-I$\alpha$ resonance line ($h\nu = 21.218$ eV). (h)-(l) Corresponding ARPES intensity plotted as a function of binding energy and wave vector.}
\end{figure*}

According to band structure calculations, the Fermi surface of FeTe$_{1- {\it x}}$Se$_{\it x}$ is basically similar to that of iron pnictides \cite{Subedi}. The parent antiferromagnetic and optimum superconducting regimes have been intensively studied by angle-resolved photoemission spectroscopy (ARPES), which has revealed that FeTe in the normal state has holelike Fermi surfaces at the $\Gamma$ point and electronlike Fermi surfaces at the M point [Fig. 1(b)].  These Fermi surfaces are connected by $Q=(\pi,0)$ similarly to the case of iron pnictides \cite{Hasan}. Furthermore, complex reconstruction of the electronic structure across the antiferromagnetic transition temperature (${\it T}_{\rm N}$) has been reported \cite{FengFeTe, Shen}. In the superconducting samples with a nearly optimum ${\it T}_{\it c}$ value, the opening of nodeless superconducting gap has also been clarified \cite{Miao, Okazaki, Nakayama}. The observed Fermi-surface dependence of the superconducting gap has been interpreted as a signature of \textit{s}$_{\pm}$-wave superconducting pairing \cite{Miao}, consistent with the appearance of so-called magnetic resonance \cite{Qiu, Mook} and the peculiar magnetic-field dependence of Bogoliubov quasiparticle interference \cite{Hanaguri}, both of which are expected for a superconducting-gap function with a sign reversal. These ARPES studies gave important insights into the origin of the antiferromagnetic and superconducting orders. However, due to the lack of systematic study on the Se concentration dependence, some essential questions remain unresolved, such as (i) how the electronic states evolve from the parent antiferromagnet to the optimum superconductor and (ii) which characteristic of the electronic states triggers superconductivity. Clarifying the low-energy electronic states as a function of Se concentration is thus of particular importance to answer these essential questions as well as to gain further insights into the superconducting mechanism.

In this Rapid Communication, we report high-resolution APRES results on iron chalcogenide FeTe$_{1- {\it x}}$Se$_{\it x}$ with various ${\it x}$  values (${\it x}  = 0, 0.2, 0.3, 0.4$, and 0.45) to show the evolution of normal-state electronic states as a function of ${\it x}$. Our systematic ARPES study definitely demonstrates that the shape of Fermi surface is almost unchanged from optimum superconducting regime ($x$ = 0.45) to non-bulk-superconducting regime ($x$ = 0.2), in sharp contrast to the case of iron pnictides \cite{Sato, Sekiba, Brouet, Kaminski, Nan, FengAsP, Cui, Feng111}. In addition, we found a significant spectral broadening in the vicinity of the antiferromagnetic phase, which characterizes an unusual metallic state at low Se concentrations. We compare the present results with the reported physical properties and discuss the implications to the origin of the characteristic superconducting phase diagram.

High-quality single-crystals of FeTe$_{1- {\it x}}$Se$_{\it x}$ were grown by the unidirectional solidification method.  The nominal compositions are FeTe (${\it x}  = 0$; ${\it T}_{\rm N}  = 67$ K), FeTe$_{0.8}$Se$_{0.2}$ (${\it x}  = 0.2$), FeTe$_{0.7}$Se$_{0.3}$ (${\it x}  = 0.3$; ${\it T}_{\it c}  = 13$ K), FeTe$_{0.6}$Se$_{0.4}$ (${\it x}  = 0.4$; ${\it T}_{\it c}  = 14$ K), and FeTe$_{0.55}$Se$_{0.45}$ (${\it x}  = 0.45$; ${\it T}_{\it c}  = 14.5$ K). Energy-dispersive x-ray spectroscopy shows that the amount of excess iron atom residing at interstitial sites is as low as 0.03. High-resolution ARPES measurements were performed with a VG-Scienta SES2002 spectrometer and a He discharge lamp ($h\nu= 21.218$ eV) at Tohoku University. ARPES measurements were also performed with synchrotron radiation at BL-28A at Photon Factory (KEK) using a VG-Scienta SES2002 spectrometer with circularly polarized 44 eV photons, and at BL-7U at UVSOR using a MBS-A1 spectrometer with linearly polarized 21 eV photons. The energy and angular resolutions were set at 6-12 meV and 0.2$^{\circ}$, respectively. Clean sample surfaces were obtained for the ARPES measurements by cleaving crystals \textit{in-situ} in an ultrahigh vacuum of 1$\times 10^{-10}$ Torr. The Fermi level ($E_{\rm F}$) of the samples was referenced to that of a gold film evaporated onto the sample holder. In this study, we adopt the one-Fe/unit-cell description where $Q=(\pi,0)$ corresponds to $Q=(\pi,\pi)$ of the two-Fe/unit-cell description adopted in our earlier study \cite{Nakayama}.

\begin{figure*}
\includegraphics[width=6.4in]{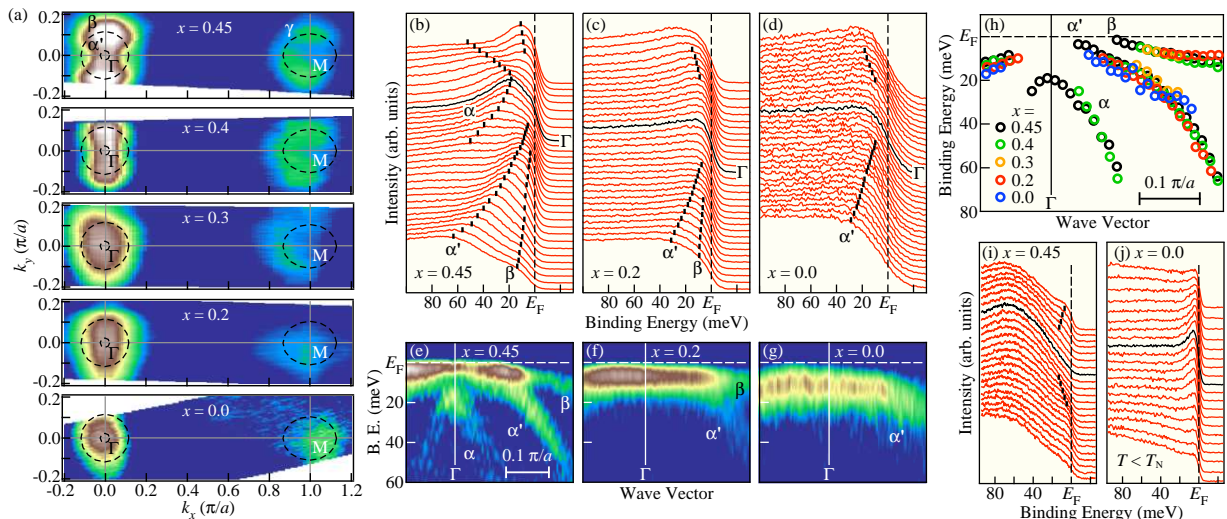}
 \vspace{-0.5cm}
 \caption{(Color online) (a) Comparison of ARPES-intensity plot at $E_{\rm F}$ as a function of two-dimensional wave vector in the normal state. The intensity is obtained by integrating the spectral intensity within $\pm10$ meV with respect to $E_{\rm F}$. The experimental Fermi surface of ${\it x} = 0.45$ (black dashed curves) determined by smoothly connecting the $k_{\rm F}$ points is superimposed on each figure for comparison. (b)-(d) High-resolution ARPES spectra near $E_{\rm F}$ and (e)-(g) corresponding second-derivative intensity plots around the $\Gamma$ point at ${\it T} = 25$ K for ${\it x} = 0.45$ and 0.2, and 80 K for ${\it x} = 0.0$. (h) Comparison of experimental band dispersions in the vicinity of $E_{\rm F}$ determined by tracing the peak position in ARPES spectra. (i) Near-$E_{\rm F}$ ARPES spectra measured around the M point at 25 K. (j) High-resolution ARPES spectra near the $\Gamma$ point obtained in the bicollinear antiferromagnetic phase (20 K) in FeTe (${\it T}_{\text N} = 67$ K). A shallow electron pocket induced by the reconstruction of the electronic structure due to the antiferromagnetic transition is observed.  Black dots in (b)-(d) and (i) are guides for eyes to trace the band dispersions.}
\end{figure*}

Figures 1(c)-1(g) compare the normal-state ARPES spectra along the $\Gamma$-M line [Fig. 1(b)] for different Se compositions. The corresponding ARPES-intensity plots as a function of binding energy and wave vector are displayed in Figs. 1(h)-1(l). As most clearly seen in the highest-${\it T}_{\it c}$  sample (${\it x} = 0.45$), there are holelike bands approaching $E_{\rm F}$ in addition to a dispersive prominent band at $\sim$300 meV binding energy near the $\Gamma$ point. We also find a less dispersive band at $\sim$50 meV and a weak but finite intensity near $E_{\rm F}$ around the M point. The latter weak intensity signifies the presence of an electron pocket as we demonstrate later. While the band structure of the five compositions is qualitatively similar, we definitely recognize the broadening of the peak structure and the reduction of intensity with decreasing ${\it x}$. We will return to this point later.

To investigate in more detail the low-energy electronic states directly responsible for superconductivity, we focus on a narrower energy range in the vicinity of $E_{\rm F}$. The ARPES intensity at $E_{\rm F}$ for different compositions is displayed in Fig. 2(a). We clearly identify an essentially similar intensity-distribution pattern for different ${\it x}$, with bright intensity around the $\Gamma$ point and relatively weak intensity around the M point, which correspond to the hole and electron Fermi surfaces, respectively. The near-$E_{\rm F}$ band dispersions around the $\Gamma$ point are composed of three holelike bands labeled $\alpha, \alpha '$, and $\beta$ [Figs. 2(b)-2(g)], which are assigned to the even combination of the $d_{xz}$ and $d_{yz}$ orbitals, the odd combination of the $d_{xz}$ and $d_{yz}$ orbitals, and the $d_{xy}$ orbital, respectively \cite{Miao}. At ${\it x} = 0.45$, the $\alpha '$ and  $\beta$  bands create hole Fermi surfaces, as seen in Fig. 2(a). While these spectral features of bands become less clear with decreasing ${\it x}$, we can still trace the dispersions of the $\alpha '$ and $ \beta$ bands until the compositions of ${\it x}  = 0.0$ and 0.2, respectively. The extracted data set in Fig. 2(h) reveals that the holelike bands crossing/touching $E_{\rm F}$ do not exhibit significant ${\it x}$ dependence (note that the $\alpha$ band is not clearly resolved for $x \leq$ 0.3, probably because of the strong spectral broadening and/or a change in the band energy).

\begin{figure}
\includegraphics[width=3.2in]{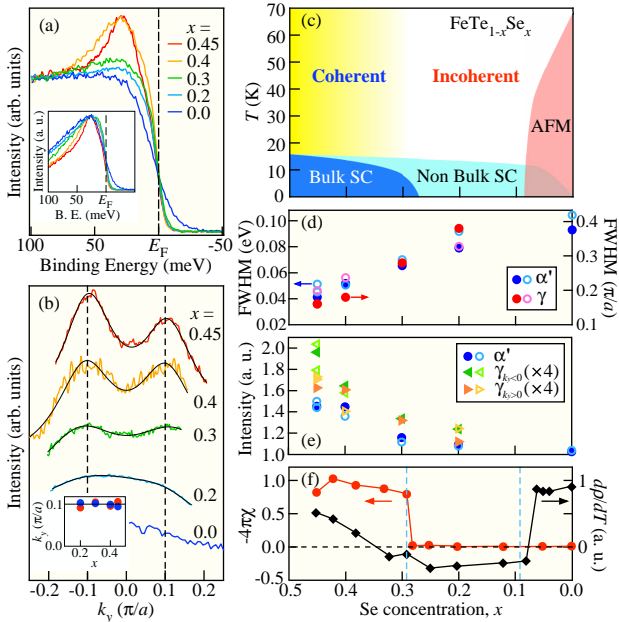}
 \vspace{-0.5cm}
 \caption{(Color online)  (a) Se-concentration dependence of the near-$E_{\rm F}$ ARPES spectrum at 20 K for $x$ = 0.2-0.45 and 80 K for $x$ = 0.0 measured at $k_y \sim 0.2$ $\pi/a$ where the quasiparticle-peak intensity of the $\alpha '$ band is dominant (note that we did not select the ARPES spectrum at the $k_{\rm F}$ point since it is strongly influenced by the $\alpha$ band).  Each spectrum is normalized to the intensity at 100 meV. The ARPES spectra after subtracting a Shirley-type background are plotted in the inset, in which each spectrum is normalized to the intensity of the peak maximum to demonstrate the spectral broadening. (b) MDCs at $E_{\rm F}$ along the $(\pi, 0)$-$(\pi,\pi)$ line for various Se concentrations.  Solid curves represent the numerical fittings with two Lorentzians. MDC peak positions extracted from the fitting are shown in the inset, where blue and red circles denote the peak position for negative and positive ${\it k_y}$ regions, respectively. (c) Phase diagram of Fe$_{1.02}$Te$_{1-{\it x}}$Se$_{\it x}$ \cite{LiuNatureMat}. (d) Full width at half maximum of the EDC peak in the inset to (a) (blue filled circle) and the MDC peak in (b) (red filled circle) plotted as a function of ${\it x}$. We found that sample-to-sample variation is much smaller than the observed $x$ dependence as highlighted by the comparison of open and filled circles. (e) ${\it x}$ dependence of the quasiparticle-peak intensity of the $\alpha '$ band (blue filled circle) extracted from ARPES spectra in (a). The MDC peak intensity extracted from (b) is also plotted using green filled triangles (${\it k_y} < 0$) and orange filled triangles (${\it k_y} > 0$). The sample-to-sample variation is also included by open symbols. (f) The superconducting volume fraction (red circle) together with the $d\rho/dT$ value at 35 K \cite{LiuNatureMat}.}
\end{figure}

Around the M point, a shallow electron pocket ($\gamma$) is observed at ${\it x} = 0.45$ [Fig. 2(i)]. Since the intensity of the $\gamma$ band is already weak at ${\it x}  =$ 0.45, its dispersion is rapidly smeared out with the reduction of Se concentration (not shown). Nevertheless, the momentum distribution curve (MDC) at $E_{\rm F}$ exhibits a peak structure showing that the $\gamma$ band resides near $E_{\rm F}$ even at ${\it x} = 0.2$ [see Fig. 3(b)].  The peak position ($k_y \sim  0.1$ $\pi/{\it a}$) corresponding to the Fermi wave vector ($k_{\rm F}$) is unchanged for $0.2$ $\leq$ ${\it x}$ $\leq$ $0.45$, although the $k_{\rm F}$  position of the $\gamma$ band is no longer well defined at ${\it x}  = 0$, as in a previous study \cite{FengFeTe}. This result, combined with the ${\it x}$-insensitive behavior of the $\alpha '$ and $\beta$ bands near the $\Gamma$ point, suggests that the shape of the Fermi surface is approximately independent for $0.2$ $\leq$ ${\it x}$ $\leq$ $0.45$, even when the Se concentration is altered from the bulk-superconducting regime (e.g., ${\it x} = 0.45$) to the non-bulk-superconducting regime (${\it x} = 0.2$).

The observed invariable Fermi-surface shape is a unique property of FeTe$_{1- {\it x}}$Se$_{\it x}$, in contrast to chemically-doped and isovalent-substituted iron pnictides, where a systematic and drastic change in the Fermi-surface size and topology takes place \cite{Sato, Sekiba, Brouet, Kaminski, Nan, FengAsP, Cui, Feng111}. In the context of the weak-coupling approach, ill-defined nesting of Fermi surface causes the disappearance of superconductivity. However, the present result strongly suggests that the ${\it T}_{\it c}$ value is not linked to the nesting condition in FeTe$_{1- {\it x}}$Se$_{\it x}$ because similar Fermi-surface nesting conditions are retained in both bulk-superconducting and non-bulk-superconducting compositions. By taking into account the absence of a good Fermi-surface nesting via $Q$ = ($\pi$, 0) in $A_x$Fe$_{2-y}$Se$_2$ \cite{Qian} and monolayer FeSe thin film \cite{Zhou} regardless of their high-${\it T}_{\it c}$ values, there would exist no common relationship between ${\it T}_{\it c}$ and the nesting condition in iron-chalcogenide superconductors. This would suggest that the Fermi-surface nesting is not important for superconductivity in these systems, while we cannot completely exclude the possible role of nesting in FeTe$_{1- {\it x}}$Se$_{\it x}$ because of the good correspondence between the development of $Q$ = ($\pi$, 0) antiferromagnetic fluctuations and the emergence of superconductivity \cite{LiuNatureMat} in the well-nested samples \cite{Nakayama}. In either case (the Fermi-surface nesting is important or not), the present results indicate that the presence of a factor which is not directly associated with the nesting should be taken into account to understand the suppression of superconductivity at low Se concentrations.

To answer the above question and further evaluate the electronic states, we turn our attention to the ARPES spectral line shape, which exhibits a striking ${\it x}$ dependence in contrast to the Fermi-surface shape.  Figure 3(a) compares the normal-state ARPES spectrum near the $\Gamma$ point for various ${\it x}$ values. While a well-defined quasiparticle peak due to the $\alpha '$ band is observed for ${\it x}$ $\geq$ $0.4$, the peak intensity is drastically suppressed and simultaneously broadened with decreasing ${\it x}$ [see inset to Fig. 3(a)], and finally almost vanishes at ${\it x} = 0$, suggesting that the electronic states become incoherent at ${\it x} = 0$ \cite{Valla}. A similar quasiparticle-peak suppression is also recognized for the electron band around the M point.  One can clearly find in Fig. 3(b) that the MDC peak around the M point is significantly broadened at the low Se concentrations. The present results thus clearly show the occurrence of progressive evolution from coherent to incoherent electronic states taking place with decreasing ${\it x}$. We have confirmed the reproducibility of such an observation by repeating the measurements at least twice for each concentration $x$.

Next we discuss the relationship between the evolution of coherent/incoherent electronic states and the disappearance of superconductivity.  In Figs. 3(c)-3(f), we show the phase diagram of FeTe$_{1- {\it x}}$Se$_{\it x}$ compared with several characteristic quantities extracted from the present ARPES and previous studies \cite{LiuNatureMat}. We immediately notice that the increase of the Se concentration gives rise to a marked sharpening of the quasiparticle-peak width [Fig. 3(d)] and an increase of the quasiparticle-peak intensity [Fig. 3(e)] as explained above. Interestingly, the ${\it x}$ dependence of the quasiparticle-peak intensity resembles the bulk superconducting-dome shape, suggesting that the development of well-defined quasiparticles in the normal state triggers bulk superconductivity in FeTe$_{1- {\it x}}$Se$_{\it x}$.  The present results agree well with the transport and superconducting properties, as explained below. As seen in Fig. 3(f), the resistivity at 35 K (slightly above ${\it T}_{\it c}$) exhibits a metallic behavior ($d \rho/ dT > 0$) for ${\it x} > 0.3$ [see black diamond in Fig. 3(f)], whereas the negative $d\rho / dT$ value for $0.1 < {\it x} < 0.3$ signifies weak charge carrier localization. For ${\it x} < 0.1$, the positive $d\rho /dT$ value is associated with the occurrence of the antiferromagnetic transition, and the charge carriers are still weakly localized ($d\rho /dT < 0$) above ${\it T}_{\rm N}$, as suggested by the absence of a Drude peak in the optical conductivity spectra \cite{Chen}. These characteristic physical properties are consistent with the observed evolution of coherent/incoherent electronic states in the normal state. We also note that a sharp quasiparticle peak expected for the metallic transport in the antiferromagnetic phase in FeTe is clearly identified around the $\Gamma$ point below ${\it T}_{\rm N}$  [Fig. 2(j)]. The good agreement of the ARPES result with the transport properties demonstrates that the observed electronic states, including the incoherent feature in the normal state, certainly reflect the inherent characteristics of FeTe$_{1- {\it x}}$Se$_{\it x}$. Most importantly, the superconducting volume fraction, which is close to 100 \% ($-4\pi \chi \sim 1$) at ${\it x} \sim 0.45$ [red circle in Fig. 3(f)], gradually decreases and eventually vanishes to $\sim$$0$ \% across the Se concentration of $\sim$$0.3$, at which the $d \rho/ dT$ value changes its sign. This demonstrates the close relationship between the emergence of incoherent normal states and the destruction of bulk superconductivity at low Se concentrations.

The origin of the incoherent electronic states at low Se concentrations is crucial for fully understanding the superconducting mechanism.  Excess iron is known to promote the carrier localization and suppress the bulk superconductivity, and thus it may be responsible for the incoherent states; e.g., Fe$_{1.03}$Te$_{0.63}$Se$_{0.37}$ is a bulk superconductor while Fe$_{1.11}$Te$_{0.64}$Se$_{0.36}$ is not \cite{LiuPRB, Sales}.  However, this would not be the main cause of the present observation, because the ARPES spectrum becomes drastically broadened even when the amount of excess iron is as low as 0.03 [see Figs. 3(a) and 3(b)]. We thus propose that either bicollinear antiferromagnetic fluctuations or electronic correlations are responsible for the incoherent states. In the former case, the bicollinear antiferromagnetic fluctuations strongly scatter itinerant electrons to make them weakly localized.  In the latter case, the large on-site Coulomb interaction pushes the system close to the Mott metal-insulator transition. The electronic correlations driven by the Hund's rule coupling may play an additional important role like in manganites \cite{ZPYin, WGYin, Ang}.

In conclusion, we reported high-resolution ARPES results of FeTe$_{1- {\it x}}$Se$_{\it x}$ with various Se concentrations ($0$ $\leq$ ${\it x}$ $\leq$ $0.45$). We found that the Fermi-surface shape does not show a clear difference for $0.2 \leq x \leq 0.45$. We also observed that the broad normal-state ARPES spectrum characterized by a small quasiparticle weight in FeTe progressively transforms into a sharp well-defined quasiparticle peak upon increasing the Se concentration. The present results suggest that the suppression of superconductivity at low Se concentrations is not caused by the deterioration of the nesting condition, but is rather associated with the incoherent electronic states produced by an additional factor like bicollinear antiferromagnetic fluctuations and electronic correlations.

\begin{acknowledgements}
We thank Y. Tanaka, H. Kumigashira, K. Ono, M. Matsunami, and S. Kimura for their assistance in ARPES measurements. This work was supported by grants from the Japan Society for the Promotion of Science (JSPS), the Ministry of Education, Culture, Sports, Science and Technology (MEXT) of Japan, the Chinese Academy of Sciences (CAS), the National Science Foundation of China (NSFC), the Ministry of Science and Technology (MOST) of China, KEK-PF (Proposal numbers: 2009S2-005 and 2012S2-001), and UVSOR (Proposal number: 22-540).
\end{acknowledgements}

\bibliographystyle{prsty}

\begin{thebibliography}{30}
 \bibitem{Kamihara}Y. Kamihara, T. Watanabe, M. Hirano, and H. Hosono, J. Am. Chem. Soc. \textbf{130}, 3296 (2008).
 \bibitem{Ren}Z. A. Ren, W. Lu, J. Yang, W. Yi, X.-L. Shen, Z.-C. Li, G.-C. Che, X.-L. Dong, L.-L. Sun, F. Zhou, and Z.-X. Zhao, Chin. Phys. Lett. \textbf{25}, 2215 (2008).
 \bibitem{Kito}H. Kito, H. Eisaki, and A. Iyo, J. Phys. Soc. Jpn. \textbf{77}, 063707 (2008).
 \bibitem{Wang}C. Wang, L. J. Li, S. Chi, Z. W. Zhu, Z. Ren, Y. Li, Y. T. Wang, X. Lin, Y. K. Luo, S. Jiang, X. F. Xu, G. H. Cao, and Z. A. Xu, Europhys. Lett. \textbf{83}, 67006 (2008).
 \bibitem{Xue}Q.-Y. Wang, Z. Li, W.-H. Zhang, Z.-C. Zhang, J.-S. Zhang, W. Li, H. Ding, Y.-B. Ou, P. Deng, K. Chang, J. Wen, C.-L. Song, K. He, J.-F. Jia, S.-H. Ji, Y.-Y. Wang, L.-L. Wang, X. Chen, X.-C. Ma, and Q.-K. Xue, Chin. Phys. Lett. \textbf{29}, 037402 (2012).
 \bibitem{Zhou}S. L. He, J. F. He, W. H. Zhang, L. Zhao, D. F. Liu, X. Liu, D. X. Mou, Y. B. Ou, Q. Y. Wang, Z. Li, L. L. Wang, Y. Y. Peng, Y. Liu, C. Y. Chen, L. Yu, G. D. Liu, X. L. Dong, J. Zhang, C. T. Chen, Z. Y. Xu, X. Chen, X. C. Ma, Q. K. Xue, and X. J. Zhou, Nature Mater. \textbf{12}, 605 (2013).
 \bibitem{Hsu}F. C. Hsu, J. Y. Luo, K. W. Yeh, T. K. Chen, T. W. Huang, P. M. Wu, Y. C. Lee, Y. L. Huang, Y. Y. Chu, D. C. Yan, and M. K. Wu, Proc. Natl. Acad. Sci. USA \textbf{105}, 14262 (2008).
 \bibitem{Yeh}K.-W. Yeh, T.-W. Huang, Y.-L. Huang, T.-K. Chen, F.-C. Hsu, P. M. Wu, Y.-C. Lee, Y.-Y. Chu, C.-L. Chen, J.-Y. Luo, D. C. Yan, and M. K. Wu, Europhys. Lett. \textbf{84}, 37002 (2008).
 \bibitem{Fang}M. H. Fang, H. M. Pham, B. Qian, T. J. Liu, E. K. Vehstedt, Y. Liu, L. Spinu, and Z. Q. Mao, Phys. Rev. B \textbf{78}, 224503 (2008).
 \bibitem{Mizuguchi}Y. Mizuguchi, F. Tomioka, S. Tsuda, T. Yamaguchi, and Y. Takano, J. Phys. Soc. Jpn. \textbf{78}, 074712 (2009).
 \bibitem{LiuNatureMat}T. J. Liu, J. Hu, B. Qian, D. Fobes, Z. Q. Mao, W. Bao, M. Reehuis, S. A. J. Kimber, K. Proke\v{s}, S. Matas, D. N. Argyriou, A. Hiess, A. Rotaru, H. Pham, L. Spinu, Y. Qiu, V. Thampy, A. T. Savici, J. A. Rodriguez, and C. Broholm, Nature Mater. \textbf{9}, 716 (2010).
 \bibitem{Li}S. Li, C. dela Cruz, Q. Huang, Y. Chen, J. W. Lynn, J. Hu, Y.-L. Huang, F.-C. Hsu, K.-W. Yeh, M.-K. Wu, and P. Dai, Phys. Rev. B \textbf{79}, 054503 (2009).
 \bibitem{MizuguchiReview}Y. Mizuguchi and Y. Takano, J. Phys. Soc. Jpn. \textbf{79}, 102001 (2010).
 \bibitem{Subedi}A. Subedi, L. Zhang, D. J. Singh, and M. H. Du, Phys. Rev. B \textbf{78}, 134514 (2008).
 \bibitem{Hasan}Y. Xia, D. Qian, L. Wray, D. Hsieh, G. F. Chen, J. L. Luo, N. L. Wang, and M. Z. Hasan, Phys. Rev. Lett. \textbf{103}, 037002 (2009).
 \bibitem{FengFeTe}Y. Zhang, F. Chen, C. He, L. X. Yang, B. P. Xie, Y. L. Xie, X. H. Chen, M. H. Fang, M. Arita, K. Shimada, H. Namatame, M. Taniguchi, J. P. Hu, and D. L. Feng, Phys. Rev. B \textbf{82}, 165113 (2010).
 \bibitem{Shen}Z. K. Liu, R.-H. He, D. H. Lu, M. Yi, Y. L. Chen, M. Hashimoto, R. G. Moore, S.-K. Mo, E.  A. Nowadnick, J. Hu, T. J. Liu, Z. Q. Mao, T. P. Devereaux, Z. Hussain, and Z.-X. Shen, Phys. Rev. Lett. \textbf{110}, 037003 (2013).
 \bibitem{Miao}H. Miao, P. Richard, Y. Tanaka, K. Nakayama, T. Qian, K. Umezawa, T. Sato, Y.-M. Xu, Y. B. Shi, N. Xu, X.-P. Wang, P. Zhang, H.-B. Yang, Z.-J. Xu, J. S. Wen, G.-D. Gu, X. Dai, J.-P. Hu, T. Takahashi, and H. Ding, Phys. Rev. B \textbf{85}, 094506 (2012).
 \bibitem{Okazaki}K. Okazaki, Y. Ito, Y. Ota, Y. Kotani, T. Shimojima, T. Kiss, S. Watanabe, C. -T. Chen, S. Niitaka, T. Hanaguri, H. Takagi, A. Chainani, and S. Shin, Phys. Rev. Lett. \textbf{109}, 237011 (2012).
 \bibitem{Nakayama}K. Nakayama, T. Sato, P. Richard, T. Kawahara, Y. Sekiba, T. Qian, G. F. Chen, J. L. Luo, N. L. Wang, H. Ding, and T. Takahashi, Phys. Rev. Lett. \textbf{105}, 197001 (2010).
 \bibitem{Qiu}Y. Qiu, W. Bao, Y. Zhao, C. Broholm, V. Stanev, Z. Tesanovic, Y. C. Gasparovic, S. Chang, J. Hu, B. Qian, M. Fang, and Z. Mao, Phys. Rev. Lett. \textbf{103}, 067008 (2009).
 \bibitem{Mook}H. A. Mook, M. D. Lumsden, A. D. Christianson, S. E. Nagler, B. C. Sales, R. Jin, M. A. McGuire, A. S. Sefat, D. Mandrus, T. Egami, and C. de la Cruz, Phys. Rev. Lett. \textbf{104}, 187002 (2010).
 \bibitem{Hanaguri}T. Hanaguri, S. Niitaka, K. Kuroki, and H. Takagi, Science \textbf{328}, 474 (2010).
 \bibitem{Sato}T. Sato, K. Nakayama, Y. Sekiba, P. Richard, Y.-M. Xu, S. Souma, T. Takahashi, G. F. Chen, J. L. Luo, N. L. Wang, and H. Ding, Phys. Rev. Lett. \textbf{103}, 047002 (2009).
 \bibitem{Sekiba}Y. Sekiba, T. Sato, K. Nakayama, K. Terashima, P. Richard, J. H. Bowen, H. Ding, Y.-M. Xu, L. J. Li, G. H. Cao, Z.-A. Xu and T Takahashi, New. J. Phys. \textbf{11}, 025020 (2009).
 \bibitem{Brouet}V. Brouet, M. Marsi, B. Mansart, A. Nicolaou, A. Taleb-Ibrahimi, P. Le F\`evre, F. Bertran, F. Rullier-Albenque, A. Forget, and D. Colson, Phys. Rev. B \textbf{80}, 165115 (2009).
 \bibitem{Kaminski}C. Liu, T. Kondo, R. M. Fernandes, A. D. Palczewski, E. D. Mun, N. Ni, A. N. Thaler, A. Bostwick, E. Rotenberg, J. Schmalian, S. L. Bud'ko, P. C. Canfield, and A. Kaminski,, Nature Phys. \textbf{6}, 419 (2010).
 \bibitem{Nan}N. Xu, T. Qian, P. Richard, Y.-B. Shi, X.-P. Wang, P. Zhang, Y.-B. Huang, Y.-M. Xu, H. Miao, G. Xu, G.-F. Xuan, W.-H. Jiao, Z.-A. Xu, G.-H. Cao, and H. Ding, Phys. Rev. B \textbf{86}, 064505 (2012).
 \bibitem{FengAsP}Z. R. Ye, Y. Zhang, F. Chen, M. Xu, Q. Q. Ge, J. Jiang, B. P. Xie, and D. L. Feng, Phys. Rev. B \textbf{86}, 035136 (2012).
 \bibitem{Cui}S. T. Cui, S. Y. Zhu, A. F. Wang, S. Kong, S. L. Ju, X. G. Luo, X. H. Chen, G. B. Zhang, and Z. Sun, Phys. Rev. B \textbf{86}, 155143 (2012).
 \bibitem{Feng111}Z. R. Ye, Y. Zhang, M. Xu, Q. Q. Ge, Q. Fan, F. Chen, J. Jiang, P. S. Wang, J. Dai, W. Yu, B. P. Xie, and D. L. Feng, arXiv:1303.0682.
 \bibitem{Qian} T. Qian, X.-P. Wang, W.-C. Jin, P. Zhang, P. Richard, G. Xu, X. Dai, Z. Fang, J.-G. Guo, X.-L. Chen, and H. Ding, Phys. Rev. Lett. \textbf{106}, 187001 (2011).
 \bibitem{Valla}T. Valla, P. D. Johnson, Z. Yusof, B. Wells, Q. Li, S. M. Loureiro, R. J. Cava, M. Mikami, Y. Mori, M. Yoshimura, and T. Sasaki, Nature(London) \textbf{417}, 627 (2002).
 \bibitem{Chen}G. F. Chen, Z. G. Chen, J. Dong, W. Z. Hu, G. Li, X. D. Zhang, P. Zheng, J. L. Luo, and N. L. Wang, Phys. Rev. B \textbf{79}, 140509(R) (2009).
 \bibitem{LiuPRB}T. J. Liu, X. Ke, B. Qian, J. Hu, D. Fobes, E. K. Vehstedt, H. Pham, J. H. Yang, M. H. Fang, L. Spinu, P. Schiffer, Y. Liu, and Z. Q. Mao, Phys. Rev. B \textbf{80}, 174509 (2009).
 \bibitem{Sales}B. C. Sales, A. S. Sefat, M. A. McGuire, R. Y. Jin, D. Mandrus, and Y. Mozharivskyj, Phys. Rev. B \textbf{79}, 094521 (2009).
 \bibitem{ZPYin}Z. P. Yin, K. Haule, and G. Kotliar, Nature Mater. \textbf{10}, 932 (2011).
 \bibitem{WGYin}W.-G. Yin, C.-C. Lee, and W. Ku, Phys. Rev. Lett. \textbf{105}, 107004 (2010).
 \bibitem{Ang}R. Ang, K. Nakayama, W.-G. Yin, T. Sato, H. Lei, C. Petrovic, and T. Takahashi, Phys. Rev. B \textbf{88}, 155102 (2013).
\end{thebibliography}

\end{document}